# Formal Probabilistic Methods for Combinatorial Structures using the Lovász Local Lemma

Chelsea Edmonds
University of Sheffield
UK
c.l.edmonds@sheffield.ac.uk

Lawrence C. Paulson
University of Cambridge
UK
lp15@cam.ac.uk

**Abstract**

Formalised libraries of combinatorial mathematics have rapidly expanded over the last five years, but few use one of the most important tools: probability. How can often intuitive probabilistic arguments on the existence of combinatorial structures, such as hypergraphs, be translated into a formal text? We present a modular framework using locales in Isabelle/HOL to formalise such probabilistic proofs, including the basic existence method and first formalisation of the Lovász local lemma, a fundamental result in probability. The formalisation focuses on general, reusable formal probabilistic lemmas for combinatorial structures, and highlights several notable gaps in typical intuitive probabilistic reasoning on paper. The applicability of the techniques is demonstrated through the formalisation of several classic lemmas on the existence of hypergraphs with certain colourings.

## 1 Introduction

In recent years there has been a surge of interest in the formalisation of mathematics using proof assistants such as Isabelle/HOL, Lean and Coq. Aside from obvious verification advantages, other benefits from the formalisation process itself include gaining new insights into proofs, the integration of advances in automation, and building up rich, searchable libraries of verified research level mathematics.

Combinatorics, once underrepresented in formal libraries, has recently seen a significant increase in formalisation efforts. Historically, the classic example was Gonthier's fundamental formalisation of the four-colour theorem [18]. More recent formalisations include our own on new libraries for combinatorial structures [16], and notable theorems from graduate or research level mathematics such as Szemerédi's regularity lemma [11, 15], the cap-set problem [10] and the Kruskal-Katona theorem [25].

Combinatorial formalisations present many interesting challenges: from the typically human intuitive nature of proofs and discrepancies in definitions, to the mixed-bag of often surprising, yet essential proof techniques from other fields. This paper explores how often intuitive probabilistic arguments on the existence of combinatorial structures translate to a formal environment. Rather than focusing on proving a singular theorem, this paper presents a more general approach to formalisation which targets proof techniques and their application in a formal environment.

The *probabilistic method* plays an increasingly important role in modern combinatorial research. The *method* refers to a vast array of probabilistic techniques and their application in a combinatorics setting, summarised by Alon and Spencer in their seminal book [3]. The basic method involves establishing a probability space over certain structures, then showing these structures have the desired properties with a positive probability. Many recent breakthroughs in combinatorics can be attributed to the application of these methods, such as Keevash's recent results on combinatorial designs [23]. Furthermore, the role of randomness in many aspects of computer science and physics, where combinatorics is often applied, has further driven the development of these techniques. This provides another motivation for their formalisation.

Despite the commonality of these methods in modern combinatorics, there are still few examples in formal libraries. The two primary examples are in Isabelle/HOL: Noschinski's girth chromatic theorem formalisation [28] and Hupel's random subgraph threshold work [21]. Alon and Spencer identify both theorems as elegant probabilistic proofs [3]. The formalisations are mostly focused on the theorems rather than reusability of the techniques. Additionally, the *dependent random choice* technique was used in Koutsoukou-Argyraki *et al.*'s formalisation of the Balog–Szemerédi–Gowers theorem [24]. While no other examples of the probabilistic method for combinatorics were found in other proof assistants, formalisations of random algorithms and several relevant basic probability concepts can be found in proof assistants such as HOL, Lean, and Coq. These are identified where relevant, and discussed in comparison to this paper in Sect. 7.

This paper will explore a number of basic and advanced techniques from the probabilistic method and demonstrate their application through proofs on hypergraph colourings. This notably includes the first formalisation of the Lovász local lemma, an important result in probability. We aim to establish a general framework through a novel use of locales, Isabelle's module system, which can be used for future formal probabilistic proofs on any type of incidence system, the basis for many combinatorial structures. We additionally significantly contribute to basic probability libraries in Isabelle. The full formalisations are available on Isabelle's Archive of Formal Proof [13, 14, 17].

In (2) we provide the necessary background, then (3) gives the formalisation of background concepts on conditional probability, independent events, and hypergraphs. Section (4) presents the basic method framework, followed by (5) which explores the formalisation of the Lovász local lemma, and finally (6) which demonstrates the application of the basic framework and Lovász local lemma to existence properties on hypergraphs. We conclude with a discussion of related work, key lessons learnt for formalising intuitive probabilistic proofs, and potential future directions in (7).

## 2 Background

### 2.1 Mathematical Background

Probability theory is built on top of the much broader field of measure theory. A *measure space* is a triplet $(X, B, \mu)$ where $(X, B)$ is a measurable space and $\mu : B \to [0, +\infty]$ is a countably additive measure. A probability space is a particularly important restricted measure space. Its definition, given below, adapts the triple's syntax to match the notation traditionally used in probability.

**Definition 2.1** (Probability Space). A *probability space* is a measure space $(\Omega, \mathcal{F}, \mathbb{P})$ which has a total measure of 1: $\mathbb{P}(\Omega) = 1$.

Commonly, $\Omega$ represents the *sample space*, the set of all possible states a random system could be in. $\mathcal{F}$ is the set of all possible events the probability space can measure, using the probability measure $\mathbb{P}$, where $\mathbb{P}(E)$ is the probability of event $E \in \mathcal{F}$ occurring. In a discrete context, $\mathcal{F} = \text{Pow}(\Omega)$. It's assumed readers have a basic knowledge of probability. A particularly important concept for this formalisation is independent events.

**Definition 2.2** (Independent events). A collection of events $E$ is defined as *independent* if and only if for all subsets $F \subseteq E$, $\mathbb{P}(\bigcap F) = \prod_{f \in F} \mathbb{P}(f)$

A related but weaker concept is that of a mutually independent set of events.

**Definition 2.3** (Mutually independent events). Given an event $A$ and a set of events $E$, $A$ is *mutually independent* of $E$ if for all subsets $F \subseteq E$, $\mathbb{P}(A \cap (\bigcap F)) = \mathbb{P}(A)\mathbb{P}(\bigcap F)$

Note that in both the above definitions we use purely set based notation, recalling that an event is simply a subset of the elements of the probability space. This notation translates directly to a formal environment compared to the intuitive logical notation typically used in mathematical text, $\mathbb{P}(A \wedge B)$.

Combinatorial applications of probability theory commonly involve discrete probability measures, which are much simpler then continuous measures. For example, discrete measures use summations rather than integrals. Most probability spaces in combinatorics involve a point measure, which assigns a specific probability to each event in the probability space. A uniform count measure is a point measure where each event has the same probability, i.e. $\mathbb{P}(A) = 1/|\Omega|$.

The core idea behind the probabilistic method for combinatorics is to show the existence of a structure with certain features by showing its probability is strictly positive. There are many techniques which can be used to obtain this positive probability bound including [3]: basic bounds; linearity of expectation; alterations; the second moment method (variance inequality); and the local lemma. More details on the basic bounds, and the local lemma will be presented alongside their formalisations.

Combinatorial structures are varied, but are often based on *incidence set systems* — a set of elements (e.g. *vertices*) and collection of subsets of those elements (e.g. *edges*). Common examples include combinatorial designs, matroids, graphs, and hypergraphs. Hypergraphs can intuitively be viewed as a generalisation of graphs where edges can be of any (non-empty) size.

### 2.2 Isabelle Background

Isabelle/HOL, henceforth referred to as Isabelle, is a proof assistant built on higher order logic. It has a number of features that make it ideal for formalising mathematics, including: the human-readable Isar proof language [33], strong automation through Sledgehammer [30], extensive foundational libraries in analysis and algebra, and the Archive of Formal Proofs (AFP) with nearly four million lines of code across entries in mathematics and computer science. As Isabelle was also used in the only prior formalisations of the probabilistic method, it is ideal to continue this type of work.

#### 2.2.1 Locales and Combinatorial Structures.
This paper builds on our previous work formalising several combinatorial structures, such as design theory [16] and graph theory [12]. These libraries use the *locale-centric* approach for formalising mathematical hierarchies, based on ideas introduced by Ballarin [5].

Locales are Isabelle's module system, enabling flexible and extensible inheritance hierarchies and proof contexts. A basic locale consists of a combination of parameters and assumptions. For example, a graph can be defined using a locale with two parameters for the vertex and edge sets, and assumptions to ensure edges are wellformed.

New locales can directly inherit from one or more existing locales, allowing for inheritance diamonds, and will often add on additional parameters or assumptions. The inheritance hierarchy can also be manipulated after locale declarations using sublocales. For example, **sublocale** $A \subseteq B$, shows that locale $A$ indirectly inherits from locale $B$.

Sublocale declarations can also use the **rewrites** command which tells Isabelle to internally rewrite all inherited facts within the current locale context using an equivalent statement for a parameter or definition.

Where possible, properties on locale structures should be declared and proved within the locale context, however, it is also often necessary to **interpret** a locale instance for use in theorem statements (global interpretation) and proofs (local interpretation). Further detail on locales is discussed in the tutorials [4] and as needed through this paper.

### 2.2.2 Existing Libraries.
Isabelle has extensive libraries in measure theory, which the probability libraries are built on. A probability space, given by Def. 2.1, is defined using the locale *prob-space*, which takes a measure space as its single parameter and an additional assumption constraining the value of the measure over the space to 1. This locale contains formal definitions for common concepts such as probability measures, expectation, variance, space and events. These are often abbreviations, i.e. pretty syntax, from concepts in measure theory and analysis. Many important lemmas are similarly inherited from the measure theory libraries.

There are many pre-defined types of measures. This paper refers to (1) the *point-measure*, which takes an additional function $p \in \Omega \to \mathbb{R}$, that "assigns a probability" to each object in the space, and (2) the *uniform-count-measure* which is a uniform specialisation of a *point-measure* where the function $p$ is not required.

## 3 Background Formalisation Work

Several significant extensions to existing libraries are required for this project, focusing on hypergraphs and probability theory. This section presents the key additions.

### 3.1 Probability

#### 3.1.1 General Event Extensions.
The *Prob-Events-Extras* theory formalised for this project contains many useful lemmas on manipulating combinations of events and calculating the resultant probabilities. This includes lemmas showing properties such as event closure and basic probability bounds on the complement, intersection and union operations, typically requiring inductive proofs, such as the example below.

**lemma** *events-inter*:
  **assumes** *finite S*  **assumes** $S \neq \{\}$
  **shows** $(\bigwedge A. \ A \in S \implies A \in events) \implies \bigcap S \in events$

Note that in Isabelle's meta-logic, $\bigwedge$, is the universal quantifier. Additionally, observe the non-empty assumption in the above proof. This is not required on paper as $\bigcap \emptyset = \mathbb{U}$, and $\mathbb{U}$ and $\Omega$ are considered interchangeable. However, this is the first sign of the *universal set vs probability space* challenge in Isabelle's probability library. In Isabelle the universal set (represented by *UNIV*) is not necessarily equal to $\Omega$.

#### 3.1.2 Conditional Probability.
There was surprisingly little support available in the existing Isabelle probability libraries for conditional probability. The most significant formalisation effort appears to be on Markov chains [20]. This introduced the *cond-prob* definition, and *cond-pmf* for working with conditional probabilities using probability mass functions (PMFs). Both are available in the main probability library; however, we found very few general lemmas.

The current notation in the *cond-prob* definition is bulky to use, so we begin our formalisation by defining an abbreviation which mirrors mathematical notation.

**abbreviation** *cond-prob-ev* :: $'a \ set \Rightarrow 'a \ set \Rightarrow real \ (\mathcal{P}'(\text{-} \mid \text{-}'))$
  **where** $\mathcal{P}(B \mid A) \equiv \mathcal{P}(x \ in \ M. \ (x \in B) \mid (x \in A))$

Conditional probability can also be viewed as the probability of $A$ given a uniform measure on $B$.

**lemma** *measure-uniform-measure-eq-cond-prob-ev3*:
  **assumes** $A \in events \ B \in events$
  **shows** $\mathcal{P}(A \mid B) = measure \ (uniform\text{-}measure \ M \ B) \ A$

This proof enables existing lemmas to effectively be lifted to the conditional probability space. Amongst the standard lemmas formalised on conditional probability, the most noticeable is *Bayes theorem* and variations. Despite its relevance and simplicity, this doesn't appear to have been previously formalised in Isabelle. However, it has been formalised in most other systems, typically as part of significant conditional probability developments, such as Mizar [31], HOL [19], Lean [32], and Coq [2, 27].

We formalise Bayes theorem first using its multiplicative form, as given below, as division can hinder automation of formal proofs. The more typical division format is also available in the final library.

**theorem** *Bayes-theorem*:
  **assumes** $A \in events \ B \in events$
  **shows** $prob \ B * \mathcal{P}(A \mid B) = prob \ A * \mathcal{P}(B \mid A)$

Particularly important to later work, is the formalisation of the general multiplication rule on events. This had not previously been formalised in Isabelle, however it does appear alongside a formalisation of Bayes theorem in other systems, such as [2].

The first challenge of this supposedly simple proof is translating the index notation commonly used on paper. There are thus two versions of this proof, one on a list of events (which imposes an ordering), and the other using a bijective indexing function on an event collection. The latter is more flexible to use, as it is relatively easy to obtain an index function given a finite collection, using bijections on sets.

**lemma** *prob-cond-inter-fn*:
  **assumes** *bij-betw g* $\{0..<card \ S\} \ S$

**assumes** *finite S* **and** $S \neq \{\}$ **and** $S \subseteq events$
**shows** *prob* $(\bigcap S) = prob\ (g\ 0) *$
  $(\prod i \in \{1..<(card\ S)\}\ .\ \mathcal{P}(g\ i\ |\ (\bigcap (g\ `\{0..<i\}))))$

Several versions of this lemma are presented in the formalisation, including specialisations involving event complements and a variation for conditional probabilities. Due to the universal set vs probability space challenge previously mentioned, in some cases the first element of the product is written separately, as shown above.

#### 3.1.3 Independent Events.
There is an existing formalisation in the main Isabelle distribution for independent sets and events. It defines *indep-event*, which takes two singular events and returns a boolean value. A more general version of this is *indep-events*, which takes a set of events and returns true if and only if they are all independent, according to Def. 2.2.

The basic *indep-event* definition had few foundational lemmas defined. While this is just a special case of the more general *indep-events* definition, the pairwise definition is useful when formalising properties on the more general definition, which often involves induction. Our formalisation contributes new introduction and elimination rules, and commutativity properties. Another useful lemma shows that if two different index functions are equivalent on the relevant set of events $E$, then one is independent if and only if the other is.

**lemma** *indep-events-fn-eq*:
  **assumes** $\bigwedge Ai.\ Ai \in E \implies F\ Ai = G\ Ai$ **and** *indep-events F E*
  **shows** *indep-events G E*

The most interesting part of this formalisation is again around event complements. Probabilistic reasoning commonly notes that switching some subset of independent events to their complements still results in an independent set without proof. The formalisation first establishes several helper lemmas on the *indep-event* pairwise definition. The main proof then proceeds by induction, but also requires a fiddly helper lemma, *indep-events-one-compl* showing that switching only one event to its complement maintained independence. The final lemma is stated below.

**lemma** *indep-events-compl*:
  **assumes** *indep-events F E* **and** *finite E*
  **shows** *indep-events* $(\lambda\ Ai.\ space\ M - F\ Ai)\ E$

Independent event formalisations can also be found in Lean's Mathlib [9], associated with significant recent work on the probability libraries such as [34]. Additionally, independent events have been formalised in Coq [27], with formalisations of independent variables also common [2].

Mutually independent events as defined in Def. 2.3 had not previously been formalised in Isabelle/HOL, or any other system to our knowledge. However, the formal definition can be stated similarly to independent events where the set $I$ represents the indexes of some set of events, and the function $F$ maps each index to an event in the probability space.

**definition** *mutual-indep-events*::  — (Definition 2.3)
  $'a\ set \Rightarrow (nat \Rightarrow 'a\ set) \Rightarrow nat\ set \Rightarrow bool$
  **where** *mutual-indep-events A F I* $\longleftrightarrow$
  $A \in events \land (F\ `I \subseteq events) \land (\forall \mathcal{J} \subseteq I\ .\ \mathcal{J} \neq \{\}) \longrightarrow$
  $prob\ (A \cap (\bigcap j \in \mathcal{J}\ .\ F\ j)) = prob\ A * prob\ (\bigcap j \in \mathcal{J}\ .\ F\ j))$

The theory contains numerous basic lemmas enabling easy reasoning on mutual independence. There are many commonalities between mutual independence and classical independence, with the latter being a stronger result. In particular, we formalise a lemma showing that a set of events $S$ is independent if and only if for every event $A \in S$, $A$ is mutually independent to the set $S \setminus \{A\}$.

**lemma** *mutual-indep-ev-set-all*:
  **assumes** $F\ `I \subseteq events$
  **assumes** $\bigwedge i.\ i \in I \implies (mutual\text{-}indep\text{-}events\ (F\ i)\ F\ (I - \{i\}))$
  **shows** *indep-events F I*

### 3.2 Hypergraphs

Hypergraphs have the same underlying foundations as combinatorial designs, which, as mentioned in Sect. 2.2, we have previously formalised [16]. Both are simply incidence set systems; however, hypergraphs are often used in different ways with their own unique concepts. For example, hypergraph language is less limited to finite structures and is more commonly used in applications of the probabilistic method.

The locale-centric approach provides an easy way to adapt the existing design theory library to mirror hypergraph language while retaining the previously proved properties. A full discussion on the hypergraph formalisation and this approach is out of scope of this paper, however, some basics required for Sect. 6 are highlighted. Prior to the design theory library work, no other proof assistants had general libraries for incidence systems, including hypergraphs.

#### 3.2.1 Designs to Hypergraphs.
We first define a *hypersystem* locale. This directly inherits from the pre-existing *incidence-set-system* locale which has two parameters representing a carrier set and a collection of subsets, defined using design theoretic language, as well as a simple well-formed assumption which ensures all vertices in edges are part of the vertex set. The **for** keyword following the direct inheritance declaration enables us to replace the prior design theoretic language used to define the parameters with hypergraph language (*vertices* and *edges*) and notation ($\mathcal{V}$ and $E$).

**locale** *hypersystem* = *incidence-system vertices* :: $'a\ set$
  *edges* :: $'a\ hyp\text{-}edge\ multiset$ **for** *vertices* $(\mathcal{V})$ **and** *edges* $(E)$

Note that $'a\ hyp\text{-}edge$ is a type synonym for $'a\ set$, and $'a\ hyp\text{-}graph$ for $'a\ set \times 'a\ hyp\text{-}edge\ multiset$. Within the locale we define numerous basic definitions such as neighbourhood, degree, adjacency and rank.

From here we continue to define different variations of hypergraphs either by direct or indirect inheritance of design concepts. For example a *hypergraph* inherits from both the *hypersystem* locale and the *inf-design* locale, which adds a non-empty edge condition. Additionally, we also formalise variations of uniform hypergraphs (constant size edges) and established inheritance with the *block-design* locale, as well as regular hypergraphs (constant degree), and establish inheritance with the *const-rep-design* locale.

These inheritances are established indirectly, as hypergraphs first define the properties in a non-finite environment. The example below demonstrates both the locale declaration for uniform hypergraphs and how to formalise this inheritance. Note that the **rewrites** command adds a proof goal, but enables us to prove the equivalence between different definitions. It then internally rewrites any inherited lemmas using the inherited definition to use the equivalent local definition within the locale context.

**locale** *kuniform-hypergraph* = *hypergraph* +
  **fixes** *k* :: *nat*
  **assumes** *uniform*: $\bigwedge e \,.\, e \in\# E \Longrightarrow card\ e = k$
**sublocale** *fin-kuniform-hypergraph-nt* ⊆ *block-design* $\mathcal{V}$ *E k*
  **rewrites** *point-replication-number E v = hdegree v*
  **and** *points-index E vs = hdegree-set vs*

### 3.2.2 Colourings.
Colourings are rarely reasoned on in design theory, but are one of the most common concepts in hypergraph (and graph) theory. As such, the hypergraph library needs to be extended to include a formalisation of hypergraph vertex colourings.

**Definition 3.1** (n-vertex colouring). A *n*-vertex colouring is an assignment of up to *n* colours to the vertices of a hypergraph such that no edge is *monochromatic*, i.e. contain vertices all the same colour.

The formalisation first defines monochromatic edges. Prior formalisations of graph colourings such as Noschinski's [29] used a simple set partition, but in this formalisation a partition approach makes it tricky to refer to an edge having a *particular* colour due to the unordered nature of sets. It also only allows for a colouring of precisely *n* colours, rather than the more general *up to n* colours in Def. 3.1, a common inconsistency in literature. As such we formalise a colouring as a function in $\mathcal{V} \rightarrow \{0.. < n\}$ where colour is a type synonym for the natural numbers.

**definition** *mono-edge* :: $('a \Rightarrow colour) \Rightarrow 'a\ hyp\text{-}edge \Rightarrow bool$
  **where** *mono-edge f e* $\equiv \exists\ c.\ \forall\ v \in e.\ f\ v = c$

The lemma *is-proper-colouring-alt* matches Def. 3.1 by unfolding the *proper-vertex-colouring* definition. The *complete-vertex-colouring* definition models a colouring using precisely *n* colours, which we show to be equivalent to the partition definition approach.

Many lemmas are available in the hypergraph library on vertex colourings. These could easily be translated to other incidence systems defined using the locale-centric approach, such as a graph theoretic context using our existing undirected graph theory library [12].

## 4 The Basic Method

The core idea behind the probabilistic method is to show the existence of a structure with certain features via a positive probability. There is a basic methodology to do this, with the calculations tending to get more complicated in line with more complex problems. This section explores the formalisation of a framework to mirror aspects of the basic method in a formal environment.

### 4.1 The Basic Method Framework
The basic method, or pattern for applying the probabilistic method on paper, can be summarised by five steps: (i) introduce randomness to the problem domain; (ii) randomly construct/select an object in the problem domain; (iii) define the desired property of this object (or property to avoid); (iv) show the desired property has a positive probability (or probability less than 1 for avoidance); (v) obtain an example of an event in the probability space with the desired property.

We propose that a 4-step formal framework can help structure formal proofs to mirror these steps.

1. Define a probability space.
2. Define object properties
3. Calculate probabilities
4. Obtain exemplar object

Note the omission of the explicit selection/construction of an object. Given the more structured way we must introduce randomness in a formal environment, most of our probability proofs are quantified over all elements of the space, so selection is done implicitly. Furthermore, while (2) is an important step, it is very problem specific so little can be done to generalise it. The remainder of this section focuses on general techniques for the remaining three steps.

### 4.2 Defining the Probability Space
Let's first look at step (1), defining a probability space. On paper, the first step introduces randomness to the problem domain in usually one informal sentence. It would be very rare that the probability space is actually defined, presenting the first challenge of formalising the probabilistic method. This framework aims to significantly simplify this step.

To establish a probability space in Isabelle, it is necessary to identify the probability measure you want to use and then interpret an instance of the *prob-space* locale in each individual proof. Additionally, to easily apply simplification tactics later in the proof, it is often useful to prove a number of additional facts around basic properties such as the space, events and measurability specific to that locale interpretation. When dealing with similar probability spaces across different proofs, this can result in notable duplication.

Noschinski's work [28] defined an *edge-space* locale, a probability space over graph edges, which introduces some modularity solving some of the above issues. Our solution significantly extends on this by taking full advantage of the flexibility of inheritance patterns with locales to develop a framework not specific to a particular measure. Firstly, we define a basic vertex space locale for probabilistic reasoning on any finite non-trivial incidence system (such as designs, graphs, or hypergraphs):

**locale** *vertex-fn-space* = *fin-hypersystem-vne* +
  **fixes** $F :: {}'a\ set \Rightarrow {}'b\ set$ **and** $p :: {}'b \Rightarrow real$
  **assumes** *ne*: $F\ \mathcal{V} \neq \{\}$ **and** *fin*: *finite* $(F\ \mathcal{V})$
  **assumes** *pgte0*: $\bigwedge fv\ .\ fv \in F\ \mathcal{V} \Longrightarrow p\ fv \geq 0$
  **assumes** *sump*: $(\sum x \in (F\ \mathcal{V})\ .\ p\ x) = 1$

Here, $F$ represents a function on the vertex set such that $\Omega = F(\mathcal{V})$ in the resultant probability space. The parameter $p$ represents a mapping from elements of $F(\mathcal{V})$ to the reals, intended to assign each element a probability. Using the *fin-hypersystem-vne* locale ensures this space could be used for practically any finite non-empty incidence system structure with at least one element in its base set.

Within the locale, two further definitions are established for notation: $\Omega = F\ \mathcal{V}$ and $M = point\text{-}measure\ \Omega\ p$. In combinatorics, which most commonly uses discrete probability spaces, a point measure is by far the most common, where some probability is assigned to each object in the space. The locale also includes formalisations of basic lemmas on the measure, finiteness properties, space, events, and measurable properties. These are all simple to formalise, but having them significantly improves automation and avoids these same basic properties being proved in each individual proof. Finally, the formalisation establishes that this locale represents a probability space, via a **sublocale** declaration. This enables probability lemmas and notation to be used naturally in the locale context.

**sublocale** *vertex-fn-space* $\subseteq$ *prob-space M*

A specialisation of a point measure is a uniform count measure, which assigns each element in the space the same probability. This is also particularly common in applications of the probabilistic method, so we define a new locale which omits the $p$ variable and defines *MU* as a *uniform-count-measure*. A simple proof establishes a **sublocale** relationship between this and the point measure counterpart.

**locale** *vertex-fn-space-uniform* = *fin-hypersystem-vne* +
  **fixes** $F :: {}'a\ set \Rightarrow {}'b\ set$
  **assumes** *ne*: $F\ \mathcal{V} \neq \{\}$ **and** *fin*: *finite* $(F\ \mathcal{V})$

**sublocale** *vertex-fn-space-uniform* $\subseteq$ *vertex-fn-space* $\mathcal{V}\ E\ F$
  $(\lambda x.\ 1\ /\ card\ \Omega U)$ **rewrites** $\Omega = \Omega U$ **and** $M = MU$

The use of **rewrites** is again particularly important, as it removes the need to unfold multiple definitions and internally rewrites the basic lemmas from the *vertex-fn-space* locale to use the parameter notation declared for the *vertex-fn-space-uniform* locale.

With these very general probability space locales in place, specialisations can be established as needed. Use of sublocales with parameters rewritten appropriately is key to this framework. The formalisation tested this process on several simple cases, such as a probability space over the vertices, or a subset of the vertex set. These could be used at the start of a proof which would naturally read something like "select a vertex at random".

In our case, we are interested in a space over a mapping from the vertex set to some property with a uniform probability distribution. This idea is formalised in the *vertex-prop-space* locale, which can be shown to be a sublocale of the *vertex-fn-space-uniform* locale given the core hypergraph parameters $\mathcal{V}$ and $E$, and the mapping which is represented by $\lambda V\ .\ V \rightarrow_E P$.

**locale** *vertex-prop-space* = *fin-hypersystem-vne* +
  **fixes** $P :: {}'b\ set$ **assumes** *finP*: *finite P* **and** *nempty-P*: $P \neq \{\}$

**sublocale** *vertex-prop-space* $\subseteq$
  *vertex-fn-space-uniform* $\mathcal{V}\ E\ (\lambda\ V\ .\ V \rightarrow_E P)$

Notably, given most of these locales build off the general *hypersystem* locale, which ultimately represents a basic incidence set system, the framework up to this point could also easily be used for other variations of set systems such as graphs and designs.

In Sect. 6, we are interested in probabilistic reasoning on random colourings of vertices in non-trivial hypergraphs. The *vertex-colour-space* locale extends a finite non-trivial hypergraph with the single parameter $n$, representing a non-zero number of colours. The formalisation also shows that it is a sublocale of the *vertex-prop-space* locale.

**locale** *vertex-colour-space* = *fin-hypergraph-nt* +
  **fixes** $n :: nat$
  **assumes** *n-lt-order*: $n \leq order$ **and** *n-not-zero*: $n \neq 0$

**sublocale** *vertex-colour-space* $\subseteq$ *vertex-prop-space* $\mathcal{V}\ E\ \{0..<n\}$
  **rewrites** $\Omega U = C^n$

Again, the **rewrites** command is integral to internally rewrite the standard definitions from the *prob-space* locale for concepts such as *space* to use the equivalent hypergraph notation, $C^n$ within the *vertex-colour-space* locale. This improves automation in later proofs and reduces the need to unfold definitions. All the basic lemmas from the original *vertex-fn-space* locale are still available, as well as other extensions from intermediate locales in this probability space hierarchy. Any proof involving a random colouring can now simply interpret this locale to set up the probability space and automatically access these properties.

This methodology naturally encourages increased modularity in proof, and thus reduces duplication. For example, general facts on vertex colouring probabilities can be formalised within the *vertex-colour-space* locale, instead of individual proofs. This is particularly valuable for lemmas that are often presented as intuitive facts on paper, but require fiddly proofs in a formal environment, that would significantly increase the proof length if included in the main proof. For example, on paper, a uniform vertex colouring could be described by saying "colour each vertex red or blue with equal probability". In the formal probability space, this actually means each vertex colouring function is equally likely. However, it would also be useful to derive a result on the probability of each individual vertex having a specific colouring, or more generally, some arbitrary property. This is a simple lemma in *vertex-prop-space*, which is automatically rewritten in *vertex-colour-space* to use $\Omega U = C^n$.

**lemma** *prob-uniform-vertex*:
  **assumes** $b \in P$ **and** $v \in \mathcal{V}$
  **shows** *prob* $\{f \in \Omega U . f v = b\} = 1/(card\ P)$

While it is intuitive that a vertex would have a colour $c$ with probability $1/n$ given $n$ colours, the formalisation requires reasoning on the cardinality of filtered sets. The *PiE-Rel-Extras* theory formalises a number of counting lemmas specific to the extensional function set relation.

### 4.3 Basic Bounds

The main task of step (2) of the framework is typically defining the *bad events* (events to be avoided), or alternatively, the desired properties of the structure. Identifying these can be a challenge in the textbook proof, but once identified should be straightforward to translate to a formal environment.

Once the properties have been identified, step (3) of the formalisation involves calculations to show the structure has the desired properties with a positive probability. These calculations can be complex, but there are a number of simple bounds which are a useful starting point. This framework formalises these basic bounds for easy applicability.

Firstly, the *union bound* intuitively states that given a collection of bad events with a total probability less than one (usually smaller), it is possible to avoid all of them [35].

**Theorem 4.1** (Union Bound). *Given events* $A = \{A_1, \ldots, A_n\}$, *then* $\mathbb{P}(\bigcup A) \leq \sum_{i=1}^{n} \mathbb{P}(E_i)$. *Therefore, if* $\sum_{i=1}^{n} \mathbb{P}(E_i) < 1$ *then* $\mathbb{P}(\overline{\bigcup A}) > 0$

The lemma *finite-measure-subadditive-finite* from the measure theory libraries previously formalised the first part of this statement in Isabelle. It is simple to extend this to show the avoidance version of the theorem for event complements.

**lemma** *Union-bound-avoid-fun*:  — (Theorem 4.1)
  **assumes** *finite A* **and** ($\sum a \in A.\ prob\ (f\ a)) < 1$ **and** $f`A \subseteq events$
  **shows** *prob* $(space\ M - \bigcup (f\ `A)) > 0$

The other bound is the *complete independence* bound [35]. Intuitively, this states that given an arbitrary number of independent bad events, each occurring with a probability less than one, then it is possible, often with a tiny probability, to avoid all of them.

**Theorem 4.2** (Complete Independence Bound). *Given a set of independent events* $A = \{A_1, \ldots, A_n\}$ *if for all* $i$, $\mathbb{P}(A_i) < 1$, *then* $\mathbb{P}(\overline{\bigcup A}) > 0$. *Note* $\overline{\bigcup A} = \bigcap_{i=1}^{n} \overline{A_i}$.

This had not previously been formalised in Isabelle, and required the lemmas on independent event complements from Sect. 3.1. The formalisation is relatively straightforward, requiring 10 Isar proof steps.

**lemma** *complete-indep-bound2-index*:  — (Theorem 4.2)
  **assumes** *finite A* **and** $F`A \subseteq events$ **and** *indep-events F A*
  **assumes** $\bigwedge a . a \in A \Longrightarrow prob\ (F\ a) < 1$
  **shows** *prob* $(space\ M - (\bigcup (F`A))) > 0$

Several versions of both bounds are available in the final library to increase applicability.

### 4.4 Obtain Structure

The final step of the framework typically obtains an exemplar object from the space with the desired property. Intuitively, this follows from demonstrating a positive probability, and is often omitted entirely from a paper proof. However, it is a necessary step in a formalisation. The framework includes the formalisation of several existence lemmas, some based on a positive probability, and the others for a probability less than one when avoiding certain events.

**lemma** *prob-lt-one-obtain*:
  **assumes** $\{e \in space\ M . Q\ e\} \in events$
  **and** *prob* $\{e \in space\ M . Q\ e\} < 1$
  **obtains** $e$ **where** $e \in space\ M$ **and** $\neg\ Q\ e$

These obtain lemmas could be easily combined with the formalisation of the union and independence bound lemmas. This effectively combines steps (3) and (4) in the formal framework and simplifies the overall proof. One example of this is given below:

**lemma** *Union-bound-obtain-fun*:
  **assumes** *finite A*
  **and** ($\sum a \in A.\ prob\ (f\ a)) < 1$ **and** $f`A \subseteq events$
  **obtains** $e$ **where** $e \in space\ M$ **and** $e \notin \bigcup \{a \in A . f\ a\}$

## 5 Lovász Local Lemma

The Lovász local lemma is a fundamental tool from the probabilistic method. It (and its variations) enable the provision of tight bounds in situations dealing with rare events, i.e. events that occur with a small positive probability. As such, it is particularly useful in step (3) of the framework. The lemma had not previously been formalised in any system. Our formalisation process begins with the general lemma, which can then be adapted to formalise several useful corollaries.

**Theorem 5.1** (General Lovász local lemma). *Let* $A_1, \ldots, A_n$ *be events in an arbitrary probability space. Suppose* $D = (V, E)$ *is a dependency (di)graph for the above events, and suppose there are real numbers* $x_1, \ldots, x_n$ *such that* $0 \leq x_i < 1$ *and* $\mathbb{P}[A_i] \leq x_i \prod_{(i,j) \in E} (1 - x_j)$ *for all* $1 \leq i \leq n$. *Then*

$$\mathbb{P}\left[\bigcap_{i=1}^{n} \overline{A_i}\right] \geq \prod_{i=1}^{n} (1 - x_i) > 0$$

Thm. 5.1 has both classical and constructive proofs available. Our formalisation follows the traditional classical proof and combines aspects of proofs from several sources, primarily including the probabilistic method textbook [3], which provides a good overview, and Zhao's probabilistic method lecture notes [35], which provided further detail.

### 5.1 Dependency Graphs

The first necessary concept for Thm. 5.1 is *dependency graphs*, a (di)graph $D = (V, E)$ where events $A_1 \ldots A_n$ are represented by $V$ and for each $i$, $1 \leq i \leq n$, the event $A_i$ is mutually independent of all the events $\{A_j : (i,j) \notin E\}$.

Interestingly, various texts will switch between using graphs and digraphs in the language. For example Zhao notes graphs are usually sufficient [35], however Alon and Spencer only reference digraphs [3]. Ultimately dependency graphs are simply an intuitive representation of mutual independence, where any events *not* in a specific event's neighbourhood are part of a mutually independent set.

As such, the formalisation could have been completed without dependency graphs. However, there can be advantages intuitively with mirroring the language used in the majority of texts, especially for formal lemmas which represent common proof techniques. Ideally, our aim is to set up the formal environment such that it is easy to switch between versions of the lemma statement with and without dependency graph notation, as done on paper.

The formalisation process quickly demonstrated that using undirected graphs would highly restrict the ability to move to a set representation. Generally, just because event $A_j$ is in a mutually independent set of $A_i$, the reverse isn't automatically true. As such, our formalisation of dependency graphs uses Noschinski's directed graph theory library [29].

**locale** *dependency-digraph* = *pair-digraph G :: nat pair-pre-digraph*
  + *prob-space M :: 'a measure* **for** *G M* +
  **fixes** $F :: nat \Rightarrow$ *'a set*
  **assumes** *vss:* $F \text{ '} (pverts\ G) \subseteq events$
  **assumes** *mis:* $\bigwedge i.\ i \in (pverts\ G) \Longrightarrow$ *mutual-indep-events*
  $(F\ i)\ F\ ((pverts\ G) - (\{i\} \cup neighborhood\ i))$

Several extensions to the library are required for this formalisation. Specifically, the original library did not include a neighbourhood definition and related basic lemmas. Additionally, we formalise a number of useful helper lemmas specific to dependency digraphs. These are derived from the mutual independence assumption, and again aim to avoid later duplication. For example, *dep-graph-indep-event* establishes an independent event set based on vertices with a zero outdegree, making use of the *mutual-indep-ev-set-all* lemma from Sect. 3.1 in its proof.

**lemma** *dep-graph-indep-events*:
  **assumes** $A \subseteq pverts\ G$ **and** $\bigwedge Ai.\ Ai \in A \Longrightarrow out\text{-}degree\ G\ Ai = 0$
  **shows** *indep-events F A*

### 5.2 Formalising the General Lemma

Using the dependency digraph library, we can now formalise Thm. 5.1 in the *prob-space* locale.

**theorem** *lovasz-local-general*:    — (Theorem 5.1)
  **assumes** $A \neq \{\}$ **and** $F\ \text{'}\ A \subseteq events$ **and** *finite A*
  **assumes** $\bigwedge Ai.\ Ai \in A \Longrightarrow f\ Ai \geq 0 \land f\ Ai < 1$
  **assumes** *dependency-digraph G M F*
  **assumes** $\bigwedge Ai.\ Ai \in A \Longrightarrow (prob\ (F\ Ai) \leq (f\ Ai) *$
  $(\prod Aj \in (pre\text{-}digraph.neighborhood\ G\ Ai).\ (1 - (f\ Aj))))$
  **assumes** *pverts G = A*
  **shows** $prob\ (\bigcap Ai \in A.\ (space\ M - (F\ Ai))) \geq$
  $(\prod Ai \in A.\ (1 - f\ Ai))$ **and** $(\prod Ai \in A.\ (1 - f\ Ai)) > 0$

There are some notable differences in the formal theorem statement. Firstly, the indices of events can be any distinct set which under an arbitrary function $F$ map to events in the probability space (rather than just $\{1, \ldots, n\}$). The function $f$ similarly maps the index set to the real numbers $x_1, \ldots, x_n$. Maintaining the indexed notation for events using $F$ is important for the lemma to be easily used. While some sources use set notation when referring to the event collection, applications of the lemma are typically to collections with no pre-existing distinctness assumptions. Lastly, *pre-digraph.neighborhood G Ai* represents the neighbourhood of vertex $Ai$ in $G$. This is an example of how local definitions from locales can still be used outside the locale context.

#### 5.2.1 The Helper Lemma.
The paper proof of the lemma focuses on a significant helper lemma containing most of the proof, which the formalisation mirrors.

**Lemma 5.2** (General Helper). *For any $S \subset 1, \ldots, n$, $|S| = s < n$ and $i \notin S$:*

$$\mathbb{P}\left[A_i \Big| \bigcap_{j \in S} \overline{A_j}\right] \leq x_i$$

This is formalised in Isabelle below, requiring all assumptions from *lovasz-local-general* except for $A \neq \{\}$, as well as introducing $S$ through new assumptions.

**lemma** *lovasz-inductive*:    — (Lemma 5.2)
  ... ⟨All *lovasz-local-general* assumptions⟩
  **assumes** *Ai-in:* $Ai \in A$ **and** *S-subset:* $S \subseteq A - \{Ai\}$
  **assumes** *S-nempty:* $S \neq \{\}$
  **assumes** *prob2:* $prob\ (\bigcap Aj \in S.\ (space\ M - (F\ Aj))) > 0$
  **shows** $\mathcal{P}((F\ Ai)\ |\ (\bigcap Aj \in S.\ (space\ M - (F\ Aj)))) \leq f\ Ai$

The proof proceeds by induction on $S$, stating the base case as trivial, before following the proof sketch below:

1. Split $S$ into $S_1 = \{j \in S | A_j \in \text{neighbourhood}(A_i)\}$, and $S_2 = S - S_1$, i.e. a set of events mutually independent of $A_i$.
2. Apply a version of Bayes rule to get the following fraction:

$$\frac{\mathbb{P}\left[A_i \cap \left(\bigcap_{j \in S_1} \overline{A_j}\right) \Big| \bigcap_{l \in S_2} \overline{A_l}\right]}{\mathbb{P}\left[\bigcap_{j \in S_1} \overline{A_j} \Big| \bigcap_{l \in S_2} \overline{A_l}\right]}$$

3. As $A_i$ is mutually independent of $S_2$, show the numerator has an upper bound of: $x_i \prod_{(i,j) \in E} (1 - x_j)$.
4. Using the induction hypothesis, show the denominator is lower bounded by: $\prod_{(i,j) \in E} (1 - x_j)$.
5. The lemma statement now follows by calculation.

The universal set vs probability space challenge again complicates the formalisation process. The textbook proof routinely uses $\mathbb{P}(\bigcap \emptyset) = \mathbb{P}(\Omega) = 1$, whereas our formalisation must deal with any probabilities conditional on $\bigcap \emptyset$ separately.

Hence, we first formalise the original base case of the lemma, showing that $\mathbb{P}(A_i) \leq x_i$ given $S = \emptyset$, in a separate lemma, *lovasz-inductive-base*. This is a straightforward formalisation requiring only four Isar proof steps, and using only four of the original general lemma assumptions.

The formalisation now proceeds with the main proof of Lemma 5.2, which first establishes some notation. The variable ?c represents a function mapping an event index to its complement event. A local instance of the digraph locale, *dg*, can also be interpreted for easy use, with an excerpt of this part of the proof below.

**interpret** *dg*: *dependency-digraph G M F*

The proof requires strong induction. Rather than inducting on the cardinality of the set $S$ as done on paper, the pre-existing *finite-psubset-induct* rule is ideal, resulting in an induction hypothesis which establishes the statement on any non-empty *proper* subset of the set $S$. Several assumptions need to be carefully selected as induction premises. The induction step of the formal proof is shown below:

**show** $\mathcal{P}((F Ai) \mid (\bigcap Aj \in S. (space M - (F Aj)))) \leq f Ai$
  **using** *finS Ai-in S-subset S-nempty prob2*
  **proof** (*induct S arbitrary*: *Ai rule*: *finite-psubset-induct* )

After applying induction, the formalisation mirrors step (1) by defining $S_1$ and $S_2$, along with a number of useful facts (finiteness, event subsets etc). Next, the formalisation shows that if $S_1 = \emptyset$, the proof follows from *lovasz-inductive-base*, as $A_i$ is mutually independent of $S_2$ by definition.

Assuming $S_1 \neq \emptyset$, steps (2) to (4) vary slightly depending on if $S_2 = \emptyset$ (due to the universal set challenge), requiring slightly different lemmas to establish the fraction and to apply the conditional multiplication rule. This case split is done on the following proof step which encapsulates the result of steps (2) to (4), to avoid duplicated work for calculations in step (5):

**moreover have** $\exists P1 P2. \mathcal{P}(F Ai \mid \bigcap Aj \in S. space M - F Aj) =$
  $P1/P2 \wedge P1 \leq prob (F Ai) \wedge P2 \geq (\prod Aj \in S1. (1 - (f Aj)))$

The cases first require slightly different lemmas to establish the fraction, as per step (2). Step (3) is straightforward in both cases in one or two formal proof steps, as it is simple to apply the mutual independence assumption. The denominator bound in step (4) which uses the induction hypothesis requires the most work. In both cases, the multiplication rule for conditional probability from Sect. 3.1 can be used. The resulting product then needs to be manipulated, as done on paper. However, typical of a formal environment, the calculations require more work. While some calculations are unique, those that are shared between cases use a single helper lemma to reduce duplication in these fiddly proofs.

From here, the formalisation completes the final calculation in step (5) using simple proof tactics.

#### 5.2.2 Applying the helper lemma.
On paper, the main lemma typically follows directly from the helper. For example, Alon and Spencer [3] state "The assertion of Lemma 5.1.1 now follows easily".

However, this isn't the case when you break the proof down formally. In particular, to apply the helper lemma, a further induction step on the events set is required. A brief survey of many lecture notes on the subject appear to routinely skip this step.

The formalisation of *lovasz-local-general* first establishes as fact the required assumptions for both the base and general version of the helper lemma, then applies the non-empty finite set induction rule with these assumptions as induction premises.

**show** *prob* $(\bigcap Ai \in A. (?c Ai)) \geq (\prod Ai \in A. (1 - f Ai))$
  **using** *assms*(3) *assms*(1) *assms*(2) *assms*(4) *general base*
  **proof** (*induct A rule*: *finite-ne-induct*)

The induction proof itself is relatively straightforward to formalise, requiring around 15 calculational Isar proof steps which use several of the lemmas on conditional probability and independence from Sect. 3.1.

### 5.3 Corollaries and Variations

There are many various forms of the Lovász local lemma. The simplest corollary states that the probability of none of the events occurring is positive. This requires a one line formalisation following immediately from the general lemma.

The symmetric Lovász local lemma is another important variation and has several forms. While less general, it is more commonly used in practice.

**Corollary 5.3** (The Lovász local lemma; symmetric case [3]). *Let $A_1, \ldots, A_n$ be events in an arbitrary probability space. Suppose that each event $A_i$ is mutually independent of a set of all the other events $A_j$ but at most $d$, and that the $\mathbb{P}[Ai] \leq p$ for all $1 \leq i \leq n$. If $ep(d+1) \leq 1$ then $\mathbb{P}\left[\bigcap_{i=1}^{n} \overline{A_i}\right] > 0$*

One commonly seen symmetric variation in literature instead retains the dependency (di)graph notation, replacing the mutually independent set condition with one that states: given a dependency graph $D = (V, E)$ where $V = \{1, \ldots, n\}$, the outdegree of each vertex is at most $d$. The second symmetric variation further replaces the $ep(d+1) \leq 1$ condition with $4pd \leq 1$, which is a tighter bound for $d < 3$.

### 5.3.1 The Symmetric Lemma; Dependency Graph.
The formalisation first proves the dependency graph representation of the theorem, as this is a more direct corollary of our earlier formalisation of Theorem 5.1.

The formal proof is split into two cases depending on whether $d = 0$. According to the textbook proof by Alon and Spencer [3], "If $d = 0$, the result is trivial". The *triviality* follows from the *dep-graph-indep-events* lemma formalised in the dependency graph locale in Sect. 5.1. From here, a positive probability can be established via the complete independence bound formalised in Sect. 4.3.

The second symmetric variation with the $4pd \leq 1$ condition only holds when $d > 0$, often skipped over on paper. For $d \geq 3$, the condition satisfies the original variation's inequality, hence the proof follows. For $d < 3$, rather than formalise a tricky inequality proof, the formalisation takes advantage of $d$ being a natural number, resulting in simple proofs for $d = 1$ and $d = 2$.

### 5.3.2 The Symmetric Lemma; Set Notation.
The symmetric lemma in its original form (Cor. 5.3) omits any reference to a dependency graph. The mutual independence condition is instead encapsulated by the following assumption: for each event $A_i$, there exists a mutually independent subset of the remaining events $A'$ such that $|A'| > |A| - d - 1$, i.e. at most $d$ other events are not in $A_i$'s mutually independent set. This single assumption replaces several assumptions on dependency graphs from the original *lovasz-local-general* theorem.

**theorem** *lovasz-local-symmetric*:    — (Corollary 5.3)
  **fixes** $d$ :: *nat*
  **assumes** $A \neq \{\}$  **and** $F \, ` \, A \subseteq events$  **and** *finite A*
  **assumes** $\bigwedge Ai. \ Ai \in A \Longrightarrow (\exists \ S \ . \ S \subseteq A - \{Ai\} \wedge$
    $card \ S \geq card \ A - d - 1 \wedge mutual\text{-}indep\text{-}events \ (F \, Ai) \ F \, S)$
  **assumes** $\bigwedge Ai. \ Ai \in A \Longrightarrow prob \ (F \, Ai) \leq p$
  **assumes** $exp(1) * p * (d + 1) \leq 1$
  **shows** $prob \ (\bigcap \ Ai \in A \ . \ (space \ M - (F \, Ai))) > 0$

This follows from *lovasz-local-symmetric-dep-graph*, requiring only two Isar proof steps. One of these steps also uses a separate helper lemma to obtain a dependency (di)graph satisfying the degree condition from the mutually independent set assumption. This obtains process is typically omitted from paper proofs, yet requires some work in a formal environment.

**lemma** *obtain-dependency-graph*:
  **assumes** $A \neq \{\}$  **and** $F \, ` \, A \subseteq events$  **and** *finite A*
  **assumes** $\bigwedge Ai. \ Ai \in A \Longrightarrow (\exists \ S \ . \ S \subseteq A - \{Ai\} \wedge$
    $card \ S \geq card \ A - d - 1 \wedge mutual\text{-}indep\text{-}events \ (F \, Ai) \ F \, S)$
  **obtains** $G$ **where** *dependency-digraph G M F* **and** *pverts G = A*
    **and** $\bigwedge Ai. \ Ai \in A \Longrightarrow out\text{-}degree \ G \ Ai \leq d$

The proof of this lemma is split into two parts. Firstly, we formalise the *define-dep-graph-set* lemma, which defines a valid dependency graph $G = (A, E)$ after acquiring a function $g$ mapping each event (vertex) to a mutually independent set using the mutual independence assumption. The second part, formalises the *define-dep-graph-deg-bound* lemma, which shows this same graph also satisfies the required out-degree condition, which follows from the assumption on the cardinality of the mutually independent set. This second condition requires some careful calculations, switching between natural numbers and integers.

The $4pd \leq 1$ variation follows a very similar proof pattern, reusing the *obtain-dependency-graph* lemma.

## 6 An Application to Hypergraph Colourings

The probabilistic method has numerous applications. The majority of techniques from the previous sections could be used to prove existence properties on numerous varieties of combinatorial structures. Hypergraph vertex colourings are a classic example. This is an ideal test case for the formal framework and bounding techniques as it is an accessible and useful area of combinatorics, as well as interesting mathematically given the duals with Ramsey theory (using edge colourings). *Property B* [3] focuses on two-colourings.

**Definition 6.1** (Hypergraph Property B [3]). A hypergraph has *Property B* if it is two-colourable, i.e. has a two-colouring where no edge is monochromatic. Let $m(n)$ denote the minimum possible number of edges of an $n$-uniform hypergraph that does *not* have *Property B*.

The probabilistic method can be used to establish existence conditions for hypergraphs which satisfy *Property B*, and in turn place bounds on $m(n)$. These properties are represented in Isabelle as follows.

**abbreviation** (**in** *hypergraph*) *has-property-B* :: *bool*
  **where** *has-property-B* $\equiv$ *is-n-colourable 2*
**definition** *min-edges-colouring* :: *nat* $\Rightarrow$ *'a itself* $\Rightarrow$ *enat*
  **where** *min-edges-colouring n -* $\equiv$
    *INF h* $\in$ (( *not-col-n-uni-hyps n*) :: *'a hyp-graph set*) .
      *enat* (*size* (*hyp-edges h*))

The *min-edges-colouring* definition uses the *INF* operator over the set of all $n$-uniform hypergraphs, denoted by the *not-col-n-uni-hyps n* definition, and returns the minimum edge size.

### 6.1 Monochromatic Edges and Independence

Basic probability properties on monochromatic edges are essential and repetitive. Building on the example in Sect. 4.2, these can be encapsulated in the *vertex-colour-space* locale. For example, we first formalise the probability of an edge $e$ being monochromatic with colour $c$ given an $n$ colouring function $f$.

**lemma** *prob-edge-colour*:
  **assumes** $e \in\# E \ c \in \{0..<n\}$
  **shows** $prob \ \{f \in C^n \ . \ mono\text{-}edge\text{-}col \ f \ e \ c\} = 1/(n \ powi \ (card \ e))$

In lecture notes, the proof of this statement is typically either glossed over [35], or mentions that as each vertex $v$ clearly independently has a colour $c$ with probability $1/n$, the independence multiplication rule can be applied.

However, this is a classic example of circular reasoning based on real world intuition when using probability. Formally, events are independent only if they adhere to the above multiplication rule. Therefore, the multiplication rule can't be used unless independence has previously been established by other logical inferences. The formalisation instead directly counts the number of colourings where an edge is monochromatic via a helper lemma on the extensional function set relation. The probability is then directly calculated using the established uniform probability rule in *vertex-fn-space*. While no longer needed, this also establishes independence on the vertex colouring events.

It is straightforward to show that the monochromatic edge event for a particular colour is disjoint from the same event for a different colour. The formalisation of the probability of an edge being monochromatic with any colour follows.

**lemma** *prob-monochromatic-edge*:
  **assumes** $e \in \# E$
  **shows** $prob\{f \in C^n \ . \ mono\text{-}edge\ f\ e\} = n\ powi\ (1 - int\ (card\ e))$

## 6.2 Property B: Uniform Hypergraphs

The following basic bound on uniform hypergraphs was proposed by Erdős in 1963. This is a classic early example of the probabilistic method on paper. The formalisation of the proof is intended to be a simple exemplar for how to apply the formal probabilistic framework from Sect. 4.

**Theorem 6.2** (Property B: $n$-uniform hypergraphs). *(i) Every $n$-uniform hypergraph with less than $2^{n-1}$ edges has property B. (ii) Therefore $m(n) \geq 2^{n-1}$.*

The proof on paper is relatively simple at approximately 5 lines.

In Isabelle, the proposition is located in the *fin-kuniform-hypergraph-nt* locale, which sets up the $n$-uniform hypergraph. Notably, this lemma does not necessarily hold if the graph is trivial — an assumption omitted from the original theorem statement. By using the framework from Sect. 4, the full formal proof requires only 11 Isar proof steps. The formalised theorem and a condensed version of the formal proof is given below.

**proposition** *erdos-propertyB*:   — (Theorem 6.2 (i))
  **assumes** *size E < ( 2^(k − 1))*   **and** $k > 0$
  **shows** *has-property-B*
 **proof** −
  **interpret** *P*: *vertex-colour-space* $\mathcal{V}$ *E 2*
    **by** *unfold-locales* (*auto simp add*: *order-ge-two*)
  **define** *A* **where** $A \equiv (\lambda\ e.\ \{f \in C^2\ .\ mono\text{-}edge\ f\ e\})$
  **have** $(\sum e \in set\text{-}mset\ E.\ P.prob\ (A\ e)) < 1$
    ⟨5-step calculation proof⟩
  **moreover have** $A\ `\ (set\text{-}mset\ E) \subseteq P.events$
    **unfolding** *A-def P.sets-eq* **by** *blast*
  **ultimately obtain** $f$ **where** $f \in C^2$ **and** $f \notin \bigcup (A\ `(set\text{-}mset\ E))$
    **using** *P.Union-bound-obtain-fun*[*of set-mset E A*] *finite-set-mset*
       *P.space-eq* **by** *auto*
  **thus** *?thesis* **using** *event-is-proper-colouring A-def*
       *is-n-colourable-def* **by** *auto*
 **qed**

The formal proof clearly lines up with each step of the formal framework as well as the original proof:

1. The first step interprets the *vertex-colour-space* locale to set up the probability space, in place of the paper proof stating "Colour $V$ randomly by two colours".
2. It then mirrors the paper proof and lets $A_e$ be the event that $e \in E$ is monochromatic (i.e. defines the event to avoid).
3. Next, the calculation step shows the sum of the probabilities of the edges being monochromatic is strictly less than one. This uses the lemma from Sect. 6.1, which the paper proof calls on without calculation. The calculations required in the 5-step Isar proof are summarised by a single line in the paper proof.
4. Finally, the *Union-bound-obtain-fun* lemma (Sect. 4.3) can be applied to obtain a colouring function not in the set of all possible monochromatic edge events (combining steps 3 and 4 of the framework).

From here it is also possible to formalise the second part of Thm. 6.2 in a few lines to establish a bound on $m(n)$.

**corollary** *erdos-propertyB-min*:   — (Theorem 6.2 (ii))
  **fixes** $z$ :: *'a itself*
  **assumes** $n > 0$
  **shows** $(min\text{-}edges\text{-}colouring\ n\ z) \geq 2^{\wedge}(n - 1)$

## 6.3 Property B: A More General Bound

Thm. 6.2 is only for $k$-uniform hypergraphs, which is a notable restriction. The Lovász local lemma enables us to establish a bound with a much more general condition.

**Theorem 6.3** (Property B). *Let $H = (V, E)$ be a hypergraph in which every edge has at least $k$ elements, and suppose that each edge of $H$ intersects at most $d$ other edges. If $e(d + 1) \leq 2^{k-1}$, then $H$ has property B.*

The proof of this property on paper begins in the same way as Thm. 6.2. There is a slight alteration to the calculation of the probability of a monochromatic edge given each edge is of a different size. It then uses two lines to establish a mutual independence condition between the different edge events, which is critical to use the Lovász local lemma. The final line of the proof simply states that the result follows from the symmetric Lovász local lemma, with no details on exactly how it is applied. The paper proof totals only 5 lines. Again, the proof also assumes a non-trivial hypergraph implicitly.

We formalised the lemma statement in the *fin-hypergraph-nt* locale, which establishes a finite non-trivial hypergraph context. The statement and step 3 of the proof is given below, clearly showing the application of the Lovász local lemma.

**proposition** *erdos-propertyB-LLL*: — (Theorem 6.3)
  **assumes** $\bigwedge e.\ e \in \#E \Longrightarrow card\ e \geq k$
  **assumes** $\bigwedge e.\ e \in \#E \Longrightarrow$
    $size\ \{\#f \in \# (E - \{\#e\#\}).\ f \cap e \neq \{\}\#\} \leq d$
  **assumes** $exp(1)*(d+1) \leq (2\ powi\ (k-1))$ **and** $k > 0$
  **shows** *has-property-B*
 **proof** –
  ⟨Framework step 1 and 2⟩
  …
  **let** *?N* = $\{0..<size\ E\}$
  **let** *?p* = $(1/(2\ powi\ (k-1)))$
  ⟨Step 3: Calculate⟩
  **have** $0 < P.prob\ (\bigcap Ai \in ?N.\ space\ P.MU - Ae\ Ai)$
  **proof** (*intro P.lovasz-local-symmetric*[*of ?N Ae d ?p* ])
   **show** $\bigwedge i.\ i \in ?N \Longrightarrow \exists S.\ (S \subseteq ?N - \{i\} \land$
    $card\ S \geq card\ ?N - d-1 \land P.mutual\text{-}indep\text{-}events\ (Ae\ i)\ Ae\ S)$
    ⟨helper lemma proof⟩
   **show** $\bigwedge i.\ i \in ?N \Longrightarrow P.prob(Ae\ i) \leq 1/(2\ powi\ (k-1))$
    **using** *P.prob-monochromatic-edge-bound*[*of - k*] ⟨proof⟩
   **show** $exp(1) * (1\ /\ 2\ powi\ int\ (k-1)) * (d+1) \leq 1$ ⟨proof⟩
  **qed** (*auto simp add*: *Ae-def E-nempty P.sets-eq P.space-eq*)
  ⟨Step 4: obtain⟩
  …
  **then show** *?thesis* **unfolding** *is-n-colourable-def* ⟨proof⟩
 **qed**

The formalisation again clearly follows the framework. Step (1) is identical, encapsulating all the shared setup between the lemmas. Step (2) similarly defines the edge event $A_e$ as before, however also uses an index function from $N$ to assign each edge a unique identifier for the Lovász local lemma, shown in the proof excerpt.

The next part of the formalisation establishes the required bound per step (3). This uses the symmetric set based Lovász local lemma from Sect. 5.3 as an introduction rule, which clearly structures the inner proof. Several of the resultant goals can be discharged automatically through existing simplification rules. This leaves three significant proof goals.

The second goal on the edge monochromatic probability inequality simply reuses the previously formalised lemma on the probability of a monochromatic edge. The third goal on the inequality, not even mentioned in the on paper proof, requires a single line automated tactic proof.

This leaves the mutually independent set condition goal. Here the formalisation diverges significantly from the original proof in [3]. Similar to the challenge on independent events, the text appears to appeal to our physical intuition to establish mutual independence, stating [3, p. 72]: "$A_e$ is clearly mutually independent of all the other events $A_f$ for all edges $f$ that do not intersect $f$." This appeared commonly across other sources, including [35], until eventually we uncovered a proof in [26] which states that this calculation follows from a theorem known as the *Mutual Independence Principle*.

No proof of this general principle is given in this source [26]. However, a proof is sketched for the lemma in this hypergraph context; stating that each event $A_e$ is mutually independent of the set of events $A_f.A_e \cap A_f = \emptyset$. Interestingly, this book specifically states: "The claim seems intuitively clear, but we should take care to prove it, as looks can often be deceiving in this field". This further motivates the formalisation.

**lemma** *disjoint-set-is-mutually-independent*:
  **assumes** *iin*: $i \in \{0..<(size\ E)\}$
  **assumes** *idffn*: $idf \in \{0..<size\ E\} \rightarrow_E set\text{-}mset\ E$
  **assumes** *Aefn*: $\bigwedge i.\ i \in \{0..<size\ E\} \Longrightarrow$
  $Ae\ i = \{f \in C^2.\ mono\text{-}edge\ f\ (idf\ i)\}$
  **shows** *prob-space.mutual-indep-events* (*uniform-count-measure* $(C^2)$) (*Ae i*) *Ae* $(\{j \in \{0..<(size\ E)\}.\ (idf\ j \cap idf\ i) = \{\}\})$

The formalisation of this lemma is over 100 Isar steps long, compared to the 15 line on paper proof sketch it is based off. The proof requires some significant manipulation of sets and variations on set filters.

Additionally, the formalisation also needs to show that the set $\{A_f.A_e \cap A_f = \emptyset\}$ meets the required size condition. This detail is entirely omitted from the proof in [3], however is comparatively a rather low effort formalisation at only 6 tidy Isar proof steps in the *intersect-empty-set-size* lemma.

Finally, an existence lemma from the final step of the basic framework can be used to obtain a proper colouring, from which the result follows.

An interesting corollary from the above statement is also included in the formalisation, showing for any $k \geq 9$, any $k$-uniform $k$-regular hypergraph $H$ has property B. This is formalised in the *erdos-propertyB-LLL9* lemma. The formalisation of this lemma requires no further probability proof steps, only a counting lemma showing an upper bound on the intersection number for an edge — a valuable addition to the hypergraph library.

## 7 Discussion

The formalisations presented in this paper offer notable insights into both the potential pitfalls of mathematical intuition in probabilistic proofs, and the challenges and advantages of formalisation at the intersection of probability and combinatorics. This section discusses key lessons learnt, bringing together themes from throughout the paper, as well as some related work.

### 7.1 Isabelle's Universal Set Challenge

A challenge that must be addressed specific to Isabelle is the disparity between the probability space, $\Omega$ and the universal set, $\mathbb{U}$. On paper, these two concepts are analogous in probability theory, which specifically enables the following calculation $\mathbb{P}(\bigcap \emptyset) = \mathbb{P}(\mathbb{U}) = \mathbb{P}(\Omega) = 1$. This is not the case in Isabelle, which made the formalisation more challenging several times throughout this work. For example, the set of all vertex colouring functions is clearly not equal to the universal set (all functions from $'a \Rightarrow nat$). Therefore, $\mathbb{U}$ contains elements outside the probability space, so $\mathbb{P}(\bigcap \emptyset) = 0$.

While possible to work around, as demonstrated in Sect. 5, the formal proofs were more complex. Another approach which could avoid this problem, while deviating from typical mathematical notation, would be to use the Isabelle PMF library. A *pmf* can be shown to inherit from the more general *prob-space* locale used in this paper. The definition also requires that $\Omega = \mathbb{U}$. However ideally we could find a solution in Isabelle for the main probability library to avoid this problem, as initial investigations indicate it is a non-issue in other proof assistants such as Lean where the probability space is identified with a type.

### 7.2 Formalising Intuition in Probability

Traditional combinatorial proof techniques such as counting rely heavily on human intuition. It was interesting to see how probability driven proofs relied on a different use of real-world intuition, often skipping over proofs of certain facts entirely. This repeatedly presented interesting formalisation challenges throughout the paper, and great opportunities to explore aspects of proofs that have not previously been looked at on paper.

A key example of this is in independence proofs where circular reasoning was surprisingly common, due to proofs that appealed to physical intuition. This intuition can perhaps be linked back to how this concept is taught early in mathematical education. For example, the Cambridge International A Level textbook [8, p.100] states "two events are said to be independent if either can occur without being affected by the occurrence of the other". It then proceeds to give the multiplication law for independent events, when in fact two events are only independent if they satisfy the multiplication law. The textbook example uses physical intuition to deduce independence, before using the multiplication law, which reinforces this circular reasoning.

In a formal setting, appealing to such physical intuition is not possible. In cases where independence was not previously established (either by calculation or assumption), the probability had to be calculated directly, which in turn required formal counting proofs. The clearest example of this was when calculating the probability of a monochromatic edge in Sect. 6.1. Mutually independent sets relied on similar physical intuition in on paper proofs. This was exemplified by the observation in Sect. 6.3 where the mutual independence principle was seldom referred to, let alone proven. This formalisation thus fills the significant gaps in the proof on paper to establish mutual independence of monochromatic non-intersecting edge events, and makes the proof easier to find to begin with.

Another interesting aspect of intuition in probability is how randomness is introduced, and results are obtained. On paper, mathematicians will usually refer to natural intuition to establish this, such as specifying individual probabilities, rather than defining the full probability space the proof is working with. This motivated the development of the general framework to structure these steps in a formal environment.

### 7.3 Reusability in Formalised Mathematics

A reoccurring challenge in formalised mathematics is the reusability of formal libraries which have been developed with a specific application in mind, and as such can be tricky to apply to other contexts or have significant gaps. Sect. 3 offers some examples of this in Isabelle in the context of conditional probability.

The methods produced in this paper aim to directly address this to minimise repeated work in formal proofs. Central to this is the framework presented in Sect. 4.1, which successfully minimises both the setup and conclusion of formal probabilistic proofs. Our approach demonstrates a new application of locales; creating a hierarchy for proof contexts rather than structures [5, 16]. Through strategic use of **rewrites**, this significantly minimised duplication between proofs on the same vertex space in Sect. 6.2 and 6.3. By basing the hierarchy on a very general incidence system locale — which is the basis of many combinatorial structures — it provides numerous exemplar formal probability space definitions which would be straightforward to apply to different types of structures in addition to hypergraphs. To test this, we refactored a probabilistic proof from prior work [24] on bipartite graphs. The framework reduced the probability space setup required, and made several lines of proof significantly simpler with a higher level of automation. This additionally reinforced the power of the locale-centric approach for mathematical hierarchies [16]. Locales were easy to use to switch between different mathematical contexts such as probability and combinatorics, and even combine ideas, as in the case of dependency graphs.

The framework is intended to be a guide for future formalisations of the probabilistic method. In addition to the probability space set up benefits, numerous variations of lemmas for the bounding and existence steps are included to make it easy to apply them naturally in different contexts. In particular, the existence lemmas to do the final step (often omitted on paper), made it easy to move from a bound to a proof conclusion. Mirroring the on paper environment, the framework enables a user to focus on the middle steps of the formalisation which are more theorem specific. The addition of several general bounding techniques to the framework, such as the Lovász local lemma, can further help structure and minimise these calculation steps, as demonstrated in Sect. 6.

### 7.4 Related Work

While there are very few prior formalisations which explore the formalisation of the probabilistic method for combinatorics [21, 24, 28], there is notable formal work exploring combinatorics and probability separately in different proof assistants. These libraries include some results formalised in Sect. 3, as previously identified.

The basic bounds presented in Sect. 4.3 are also relatively simple concepts in probability. The union bound is also known as *Boole's inequality*, which is formalised in Coq's measure theory libraries [7]. Additionally, it is explored from an entirely different angle as a basis for a program logic formalised in Coq [6]. While there are no clear past formalisations of the complete independence bound, it is possible there are similar underlying concepts formalised in the extensive Coq and HOL libraries on measure theory, that would be easy to lift to a probability space context. The differentiating factor in this paper remains the focus on reusability, particularly in the context of combinatorics. Both basic bounds were formalised several ways, enabling future formal proofs to use these properties without the need to understand the complex underlying measure theory libraries.

Beyond mathematical formalisations, several projects have explored the verification of random algorithms and probabilistic aspects of programming. Many of the Coq references in Sect. 3 were motivated by these types of projects. In Isabelle, there are several examples of probabilistic algorithms on graphs that have been very recently formalised. This includes work on the RANKING argument [1], and expander graphs [22]. The latter is more mathematical, and makes several contributions to the probability libraries which could be useful in the probabilistic method framework, however uses PMFs. An interesting avenue for future work would be to look at formalising a constructive proof of the Lovász local lemma, which would likely benefit from past formalisations of probabilistic algorithms, in comparison to the approach presented in this paper.

## 8 Concluding Comments

This paper proposed a general formal framework for proofs using the probabilistic method in combinatorics, a fascinating intersection of two mathematical fields. The framework makes it easier to translate intuitive aspects of probability proofs to the formal environment, while reducing duplication between proofs in the context of combinatorial structures based on incidence systems. A significant aspect of this framework is the first formalisation of the Lovász local lemma — a fundamental technique in probability with wide application potential — alongside other contributions to general libraries on probability and combinatorics which could be used in a wide range of future work. Exploring proofs on hypergraph colourings additionally uncovered some fascinating discrepancies in mathematical intuition in the probabilistic context. The formalisations are available on the Isabelle Archive of Formal Proofs [13, 14, 17] for easy access. The framework and all related lemmas were kept intentionally general, opening the door to future extensions such as further probabilistic methods, and new applications across different combinatorial structures.

## Acknowledgements

Thanks to the anonymous reviewers for their useful and thoughtful feedback which was incorporated into the final version of this paper.

Chelsea Edmonds was jointly funded for this work by the Cambridge Trust (Cambridge Australia Scholarship) and a Cambridge Department of Computer Science and Technology Premium Research Studentship. Lawrence C. Paulson was funded by the ERC Advanced Grant ALEXANDRIA (Project GA 742178).